\documentclass[aps,prl,reprint,showpacs,showkeys,superscriptaddress]{revtex4-1}
\usepackage{graphicx}
\usepackage{bm}
\usepackage{color}
\usepackage{amsmath}
\usepackage{amsfonts}
\usepackage{amssymb}
\usepackage{natbib}
\usepackage{latexsym}
\usepackage{float}
\usepackage{verbatim}
\usepackage{hyperref}

\begin{document}

\title{Particle Acceleration in Relativistic Plasma Turbulence}
\author{Luca Comisso}
\email{luca.comisso@columbia.edu}
\affiliation{Department of Astronomy and Columbia Astrophysics Laboratory, Columbia University, New York, NY 10027, USA}
\author{Lorenzo Sironi}
\email{lsironi@astro.columbia.edu}
\affiliation{Department of Astronomy and Columbia Astrophysics Laboratory, Columbia University, New York, NY 10027, USA}

%\date{\today}

\begin{abstract}
Due to its ubiquitous presence, turbulence is often invoked to explain the origin of nonthermal particles in astrophysical sources of high-energy emission. With particle-in-cell simulations, we study decaying turbulence in magnetically-dominated (or equivalently, ``relativistic'') pair plasmas. We find that the generation of a power-law particle energy spectrum is a generic by-product of relativistic turbulence. The power-law slope is harder for higher magnetizations and stronger turbulence levels.
In large systems, the slope attains an asymptotic, system-size-independent value, while the high-energy spectral cutoff increases linearly with system size; both the slope and the cutoff do not depend on the dimensionality of our domain. 
By following a large sample of particles, we show that particle injection happens at reconnecting current sheets; the injected particles are then further accelerated by stochastic interactions with turbulent fluctuations.
 Our results have important implications for the origin of non-thermal particles in high-energy astrophysical sources.
\end{abstract}

%\pacs{52.27.Ny; 52.30.Cv; 52.35.Vd, 04.20.-q}
%\keywords{Relativistic plasmas; Turbulence; Particle acceleration}

\maketitle
{\it Introduction.---} 
Despite decades of research \cite{fermi49,parker58,sturrock66,kulsrud71,bell78,BO78}, the origin of nonthermal particles in space and astrophysical plasmas remains poorly understood. Due to its ubiquitous presence, turbulence is often invoked as a promising source of accelerated particles \cite{melrose80,lazarian12,Petrosian12}, and significant progress has been made on both theoretical \cite{pryadko97,schlickeiser98,chandran2000,brunetti07,lynn14} and numerical \cite{matthaeus84,dmitruk04,dalena14,matsumoto15,isliker17} grounds. Turbulence is believed to play an important role in the energization of nonthermal particles in the solar corona and galaxy clusters \cite{melrose80,lazarian12,Petrosian12}, and it could also be important in magnetically-dominated environments like pulsar magnetospheres and winds, jets from active galactic nuclei, and coronae of accretion disks. Particle acceleration in magnetized turbulent flows might indeed power the bright nonthermal synchrotron and inverse Compton signatures from such high-energy sources \cite{begelman84,yuan14,buhler14}.

While the dynamics of turbulent flows in magnetically-dominated plasmas (or equivalently, in the ``relativistic'' regime in which the magnetic energy exceeds the plasma rest mass energy) has been well characterized by fluid simulations \cite{cho05,inoue11,zrake12,zrake13,cho14,zrake14,takamoto16,zrake16,takamoto17}, the process of particle acceleration can only be captured from first principles by means of fully-kinetic particle-in-cell (PIC) codes. Pioneering studies of particle acceleration via driven turbulence in moderately magnetized pair plasmas \cite{zhdankin17} reported the generic development of nonthermal power-law distributions. However, the power-law tail was found to steepen with increasing system size, with disappointing implications for large-scale astrophysical sources. Here, by employing PIC simulations in unprecedentedly large domains, we show that the power-law slope reaches an asymptotic, system-size-independent value, with harder slopes for higher magnetizations and stronger turbulence levels. By following a large sample of particles, we show that particle injection happens at reconnecting current sheets; the injected particles are then further accelerated by stochastic interactions with turbulent fluctuations.

{\it Method and setup.---} 
To study the physics of particle acceleration from first principles, we solve the coupled Vlasov-Maxwell system of equations through the PIC method \cite{birdsall85}. 
%which evolves electromagnetic fields via Maxwell's equations and  particle trajectories via the Lorentz force. 
We employ the electromagnetic fully-relativistic three-dimensional (3D) PIC code TRISTAN-MP \cite{buneman93,spitkovsky05} to perform 2D and 3D simulations of decaying  turbulence in pair plasmas. In 2D our domain is a periodic square of side $L$ in the $xy$ plane, in 3D it is a cube. The electron-positron plasma is initially uniform with density $n_0$ and follows a Maxwellian distribution with thermal spread $\theta_0 = {k_B T_0}/{m c^2} = 0.3$.
We set up a mean magnetic field $\left\langle {\bm{B}} \right\rangle  = B_0 {\bm{\hat z}}$ and magnetic field fluctuations $\delta B_x$ and $\delta B_y$, whose strength is parameterized by the magnetization $\sigma_0  = \delta B_{{\rm{rms}}0}^2/4\pi n_0 w_0 m c^2 $, where  $\delta B_{{\rm{rms}}0}^2 = \langle {\delta {B^2}} \rangle_{t=0}$, and $w_0 = \gamma_{th0}+\theta_0$ is the initial enthalpy per particle ($\gamma_{th0}$ is the mean particle Lorentz factor). We vary $\sigma_0$ from $2.5$ to $160$ (i.e., in the magnetically-dominated regime $\sigma_0\gg1$, where the Alfv{\'e}n speed approaches the speed of light) and $\delta B_{{\rm{rms}}0}/B_0$ from 0.5 to 4. With our definition of $\sigma_0$, our results do not depend on the choice of initial thermal spread $\theta_0$ (apart from an overall energy rescaling). We also define $\sigma_z=B_0^2/4\pi n_0 w_0 m c^2 =\sigma_0(B_0^2/\delta B_{{\rm{rms}}0}^2)$.

Turbulence develops from uncorrelated fluctuations with $\delta B_x = \sum\nolimits_{m,n} {\beta_{mn} n \sin ({k_m}x + \phi_{mn})} \cos ({k_n}y + \varphi_{mn})$ and $\delta B_y= - \sum\nolimits_{m,n} {\beta_{mn} m \cos ({k_m}x + \phi_{mn})} \sin ({k_n}y + \varphi_{mn})$, where $m,n \in \left\{ {1, \ldots ,N} \right\}$ are the mode numbers, $k_m=2\pi m/L$ and $k_n=2\pi n/L$ the wavenumbers along $x$ and $y$ respectively, $\phi_{mn}$ and $\varphi_{mn}$ random phases, and $\beta_{mn}=2\,\delta B_{{\rm{rms}}0}/[N(m^2+n^2)^{1/2}]$. With this choice, each $(m,n)$ mode carries the same power, so the initial energy spectrum peaks near $k_N=2\pi N/L$ (typically, $N=8$). This defines the energy-carrying scale $l=2 \pi/k_N$, used as our unit length. For 3D simulations, we also modulate $\delta B_x$ and $\delta B_y$ sinusoidally in the $z$-direction, with two modes of wavelength $L$ and $L/2$ and random phases. The initial setup is not in pressure balance and the system rapidly evolves into a turbulent state.

The large size of our computational domain (with $L$ up to 65,600 cells in 2D and up to 2400 in 3D) allows to achieve asymptotically-converged results. The plasma skin depth $d_{e0}=c/\omega_{p0}=\sqrt {\gamma_{th0} {m}c^2/4\pi n_0 {e^2}}$ is resolved with 10 cells in 2D and 3 cells in 3D (in 2D we have checked that $d_{e0} = 3$ or 10 cells give identical results). The simulation timestep is controlled by the numerical speed of light of 0.45 cells/timestep. We typically employ 16 (macro)particles per cell in 2D and 4 per cell in 3D, but we have tested that our results are the same when using more particles per cell (up to 256 in 2D and up to 16 in 3D).

\begin{figure}
\begin{center}
\hspace*{0.215cm}\includegraphics[width=8.65cm]{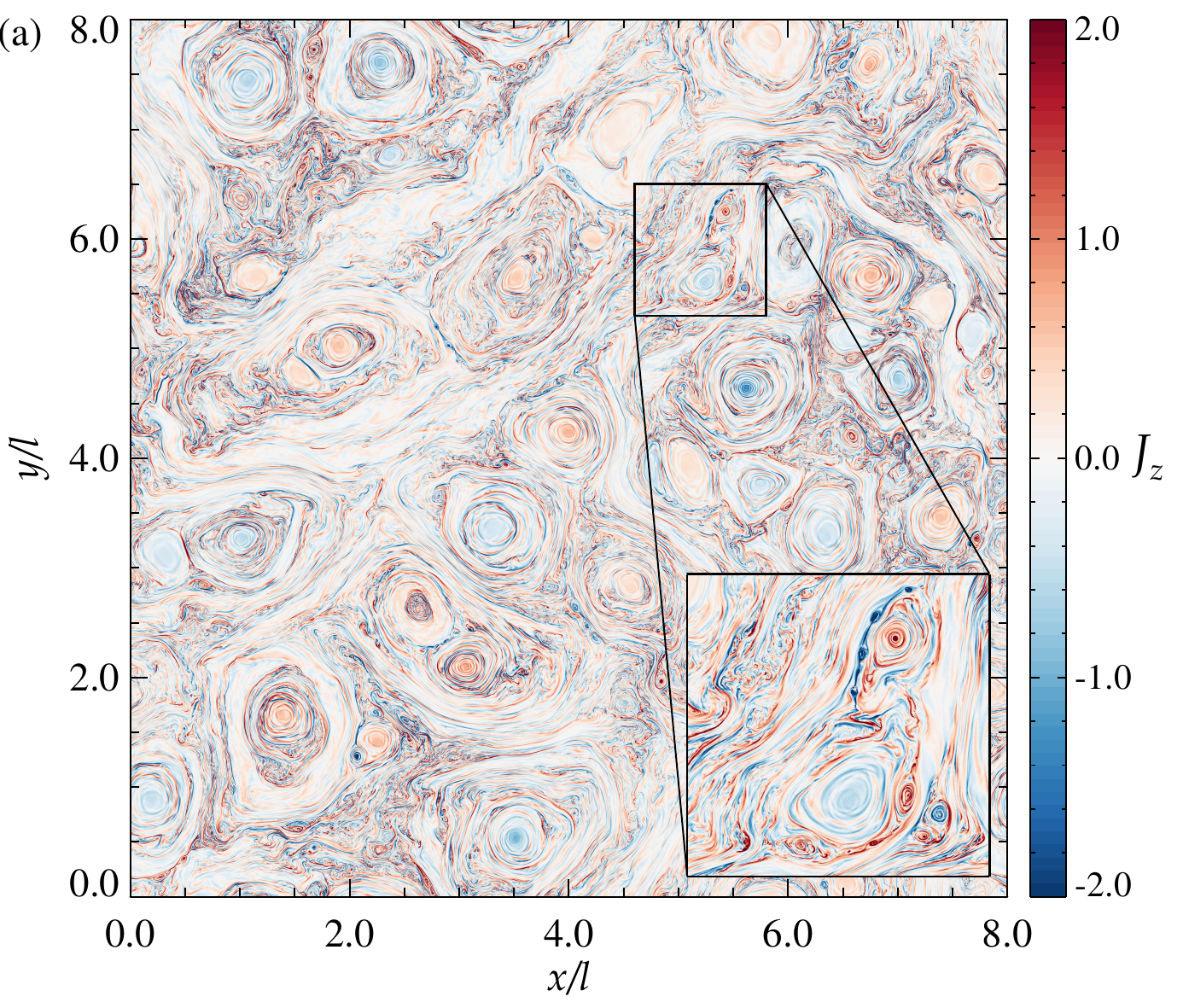}
\includegraphics[width=8.65cm]{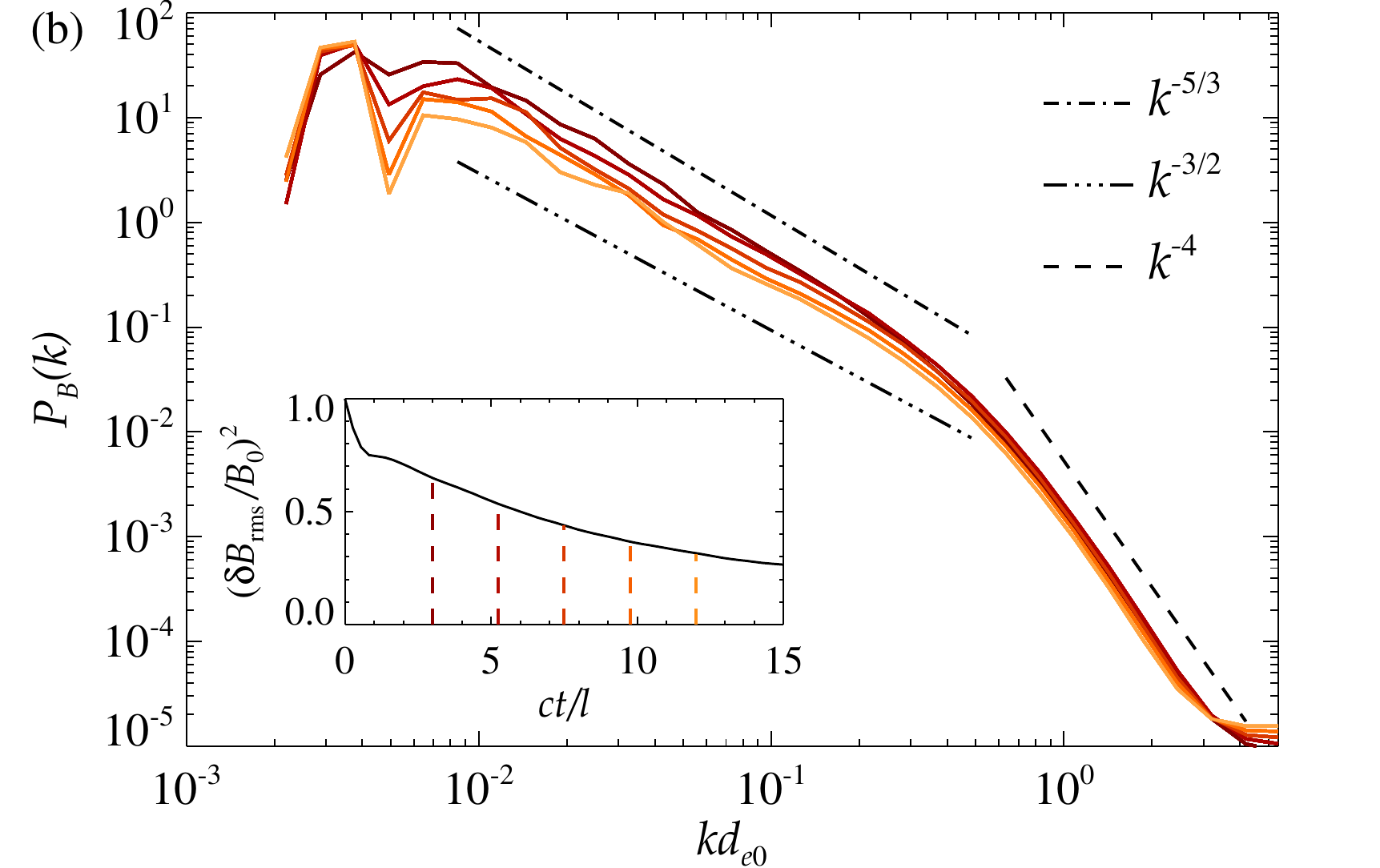}
\end{center}
\caption{Development of turbulence from a 2D simulation with $\sigma_0=10$, $\delta B_{{\rm{rms}}0}/ B_0=1$, and $L/d_{e0}=3280$ (with $l=L/8$). Top: Current density $J_z$ at $ct/l=5.5$ (normalized to $e n_0 c$)  indicating the presence of coherent structures like current sheets, plasmoids, and vortices (see inset) \cite{Movies}. Bottom: Magnetic power spectrum, showing a well-developed inertial range and a kinetic range with $P_B(k) \propto k^{-4}$ (dashed line). The inset shows the time evolution of $\delta B_{\rm rms}^2=\langle {\delta {B^2}}\rangle$ normalized to $B_0^2$, with vertical dashed lines indicating the times when the magnetic power spectra presented in the main panel are computed (same color coding).}
\label{fig1}
\end{figure}

{\it Turbulence and nonthermal particle spectrum.---} 
Fig.~\ref{fig1}(a) shows the fully-developed turbulent state from a 2D simulation with $\sigma_0=10$, by plotting the out-of-plane current density $J_z$. Vortex-like and sheet-like coherent structures are ubiquitous, in analogy to nonrelativistic kinetic simulations \citep[e.g.,][]{servidio12,wan12,tenbarge13,franci15,makwana15}. Elongated current sheets tend to fragment into a chain of plasmoids/magnetic islands, due to the 
% are present at different scales, with a lower thickness set by $\delta_{cs}  \sim \sqrt {w/(4 \pi n^2 e^2)} = \sqrt{\gamma_{th} + \theta} \, c/\omega_p$ \cite{comisso14}. 
plasmoid instability  \cite{comisso16,huang17,TajimaShibata97,loureiro07}.
%, which further mediates the transfer of energy to smaller scales . 
As we show below, reconnecting current sheets --- a natural by-product of turbulent cascades in magnetized plasmas  \cite{carbone90,mallet17,loureiro17,comisso18} --- play a vital role for particle injection into the acceleration process.
The time evolution of the magnetic power spectrum $P_B(k)$ is presented in Fig.~\ref{fig1}(b), where $P_B(k) dk = \sum \nolimits_{{\bm{k}} \in dk} {({\bm{B}}_{\bm{k}} \cdot {\bm{B}}_{\bm{k}}^*)/B_0^2}$ is computed from the discrete Fourier transform ${\bm{B}}_{\bm{k}}$ of the field. Each curve refers to a different time (from brown to orange), as indicated by the corresponding vertical dashed lines in the inset, where we present the temporal decay of the energy in turbulent fluctuations $\delta B^2_{\rm rms}/B_0^2$. As the magnetic energy decays (by the end of our simulation, $\sim 70\%$ of the initial turbulent energy has been converted to particle energy), the inertial range ($kd_{e0}\lesssim 0.4$) of the magnetic power spectrum tends to flatten from $P_B(k) \propto k^{-5/3}$ \cite{biskamp03,thompson98} to $P_B(k) \propto k^{-3/2}$ \cite{Iroshnikov63,Kraichnan65}. At kinetic scales ($kd_{e0}\gtrsim 0.4$), the spectrum steepens to $P_B(k) \propto k^{-4}$, similar to what has been found in kinetic simulations of driven turbulence with moderate magnetizations \cite{zhdankin17,zhdankin18arXiv}.

\begin{figure}[]
\begin{center}
\includegraphics[width=8.65cm]{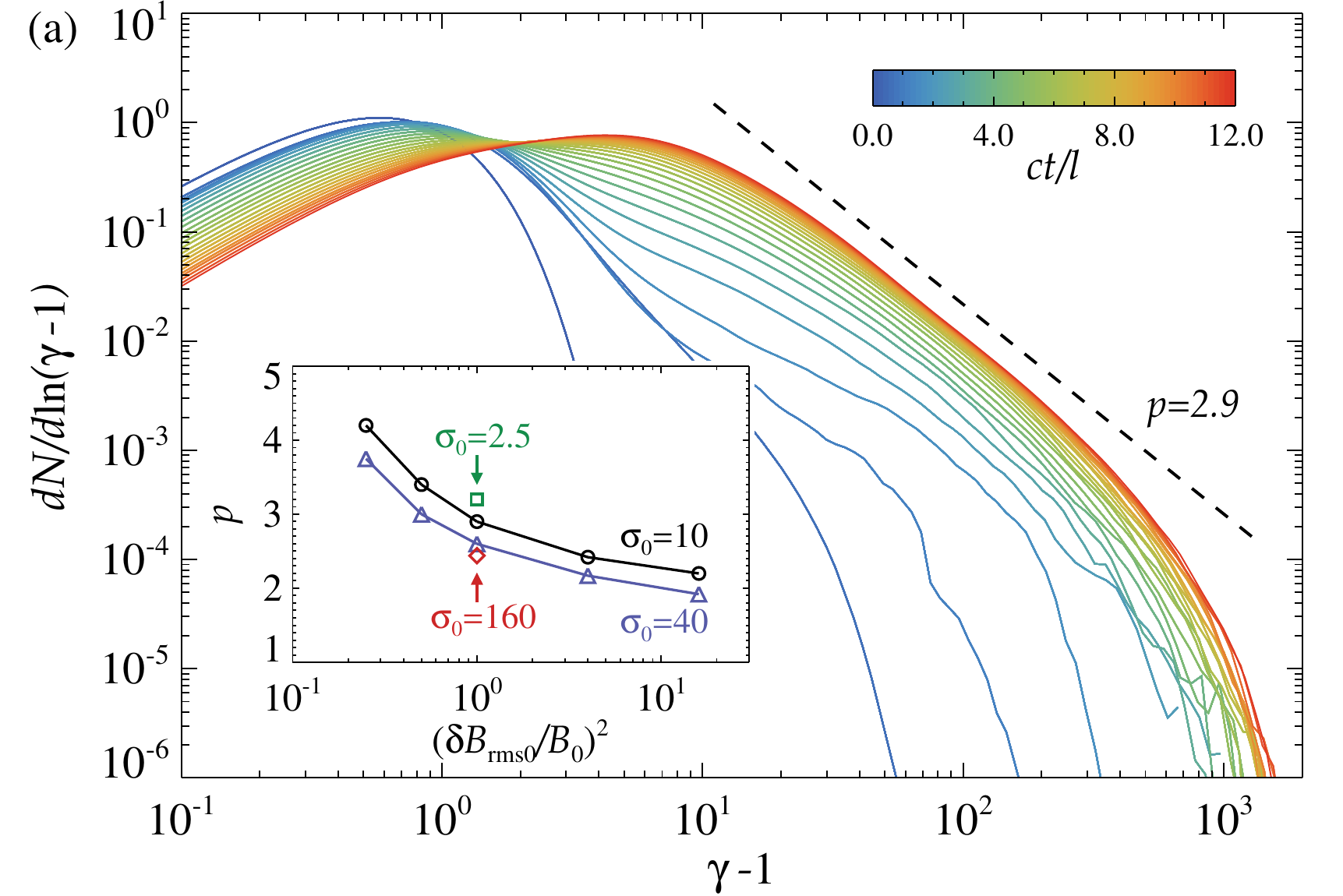}
\includegraphics[width=8.65cm]{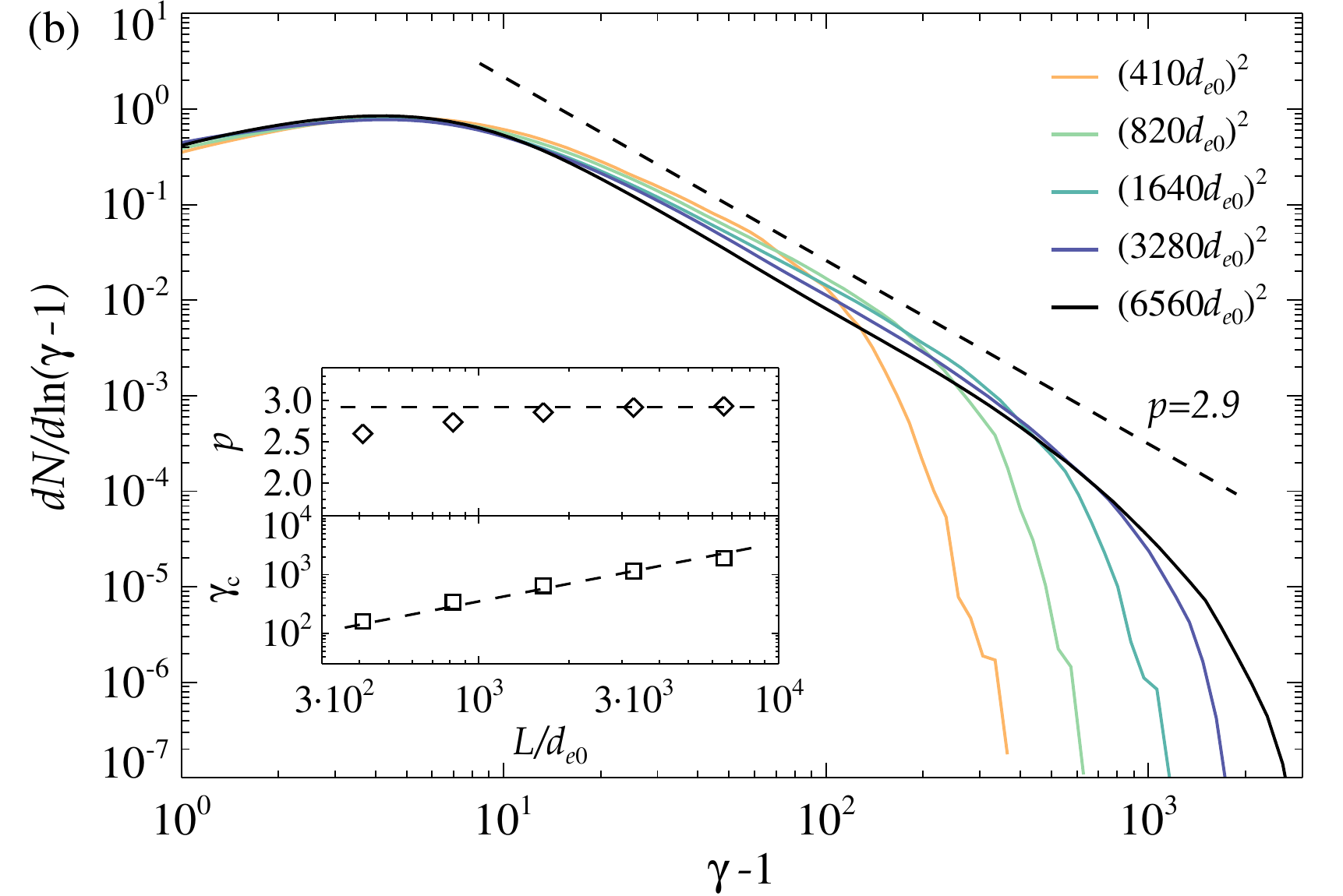}
\end{center}
\caption{Top: Time evolution of the particle spectrum for the simulation in Fig.~\ref{fig1}. At late times, the spectrum displays an extended power-law tail with slope $p = - d\log N/d\log (\gamma -1)\sim 2.9$. The inset shows the dependence of the power-law index $p$ on $\delta B_{{\rm{rms0}}}^2/B_0^2$ and $\sigma_0$. Bottom: Particle spectra at late times ($ct/l=12$) for simulations with fixed $\sigma_0=10$, $\delta B_{{\rm{rms}}0}/ B_0=1$ and $l=L/8$, but different system sizes $L/d_{e0} \in \left\{ {410,820,1640,3280,6560} \right\}$. The insets show the dependence of the power-law index $p$ (top; dashed line is the asymptotic slope $p=2.9$) and the cutoff Lorentz factor $\gamma_c$ (bottom; dashed line is the predicted scaling $\gamma_c\sim \sqrt{\sigma_z} \gamma_{th0} (l/d_{e0})$, see text) on the system size.}
\label{fig2}
\end{figure}

The time evolution of the corresponding particle spectrum $dN/d\ln(\gamma-1)$ is presented in Fig.~\ref{fig2}(a), where $\gamma$ is the particle Lorentz factor. The figure shows that efficient nonthermal particle acceleration is a self-consistent by-product of relativistic turbulence. As a result of field dissipation, the spectrum shifts to energies much larger than the initial  Maxwellian (which is shown by the blue line peaking at $\gamma-1\sim \gamma_{th0}-1\simeq 0.6$). At late times, when most of the turbulent energy has decayed, the spectrum stops evolving (orange and red lines): it peaks at  $\gamma-1\sim \gamma_{th0}(1+\sigma_0/2)-1\simeq 4$, and extends well beyond the peak into a nonthermal tail of ultra-relativistic particles, with power-law slope $p = - d\log N/d\log (\gamma-1) \sim 2.9$.
%(compare with the dashed line). 
The inset shows that the value of the power-law slope is not universal: for fixed $\delta B_{{\rm{rms0}}}^2/B_0^2$, the tail becomes harder with increasing $\sigma_0$, in agreement with \cite{zhdankin17} and in analogy to the results of PIC simulations of relativistic magnetic reconnection \cite{sironi14,guo14,werner16,lyutikov17,petropoulou18}; more dramatically, at a fixed magnetization $\sigma_0$, the spectrum is much harder for stronger turbulent fluctuations (i.e., at $\delta B_{{\rm{rms0}}}^2/B_0^2\gtrsim 1$).

The power-law slopes quoted in the inset of Fig.~\ref{fig2}(a) persist in the limit of asymptotically large domains. %\cite{pslope}. 
In Fig.~\ref{fig2}(b), we show for $\sigma_0=10$ and $\delta B_{{\rm{rms0}}}^2/B_0^2=1$ the dependence of the time-saturated particle spectrum on the size of our 2D box, which we vary in the range $L/d_{e0} \in \left\{ {410,820,1640,3280,6560} \right\}$. While earlier works, that employed smaller domains, had claimed that the power-law slope steepens with increasing system size \cite{zhdankin17}, we find that the slope saturates for asymptotically large systems (top inset in Fig.~\ref{fig2}(b)), which allows us to extrapolate our results to the astrophysically-relevant regime $L/d_{e0}\gg1$. On the other hand, the high-energy cutoff $\gamma_c$ --- defined as the Lorentz factor where the spectrum drops one order of magnitude below the power-law best fit --- linearly increases with system size (bottom inset in Fig.~\ref{fig2}(b)). As discussed below, stochastic acceleration by turbulent fluctuations dominates the energy gain of nonthermal particles. High-energy particles will cease to be efficiently scattered by turbulent fluctuations when their Larmor radius exceeds the energy-carrying scale $l=2\pi/k_N$, implying an upper limit to their Lorentz factor of $\gamma_c\sim e \sqrt{\langle B^2 \rangle} l/m c^2\sim \sqrt{\sigma_z} \gamma_{th0} (l/d_{e0})$, which successfully matches the scaling of $\gamma_c$ on system size in the inset of Fig.~\ref{fig2}(b) (this argument assumes that the turbulence survives long enough to allow the particles to reach this upper limit). 
By varying $l/L$, we have explicitly verified that $\gamma_c\propto l$, rather than $\gamma_c\propto L$.
 
\begin{figure}[]
\begin{center}
\includegraphics[width=8.65cm]{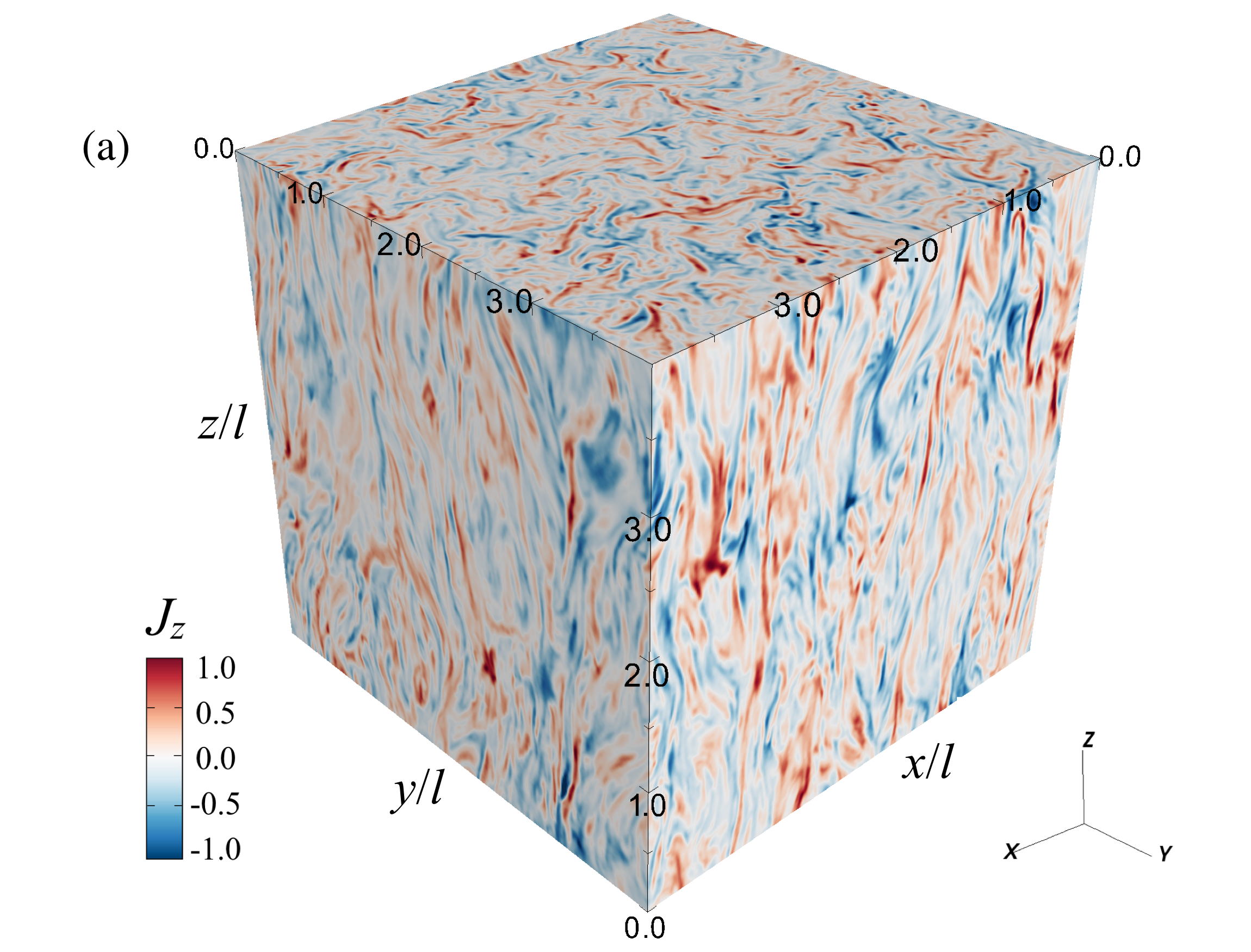}
\includegraphics[width=8.65cm]{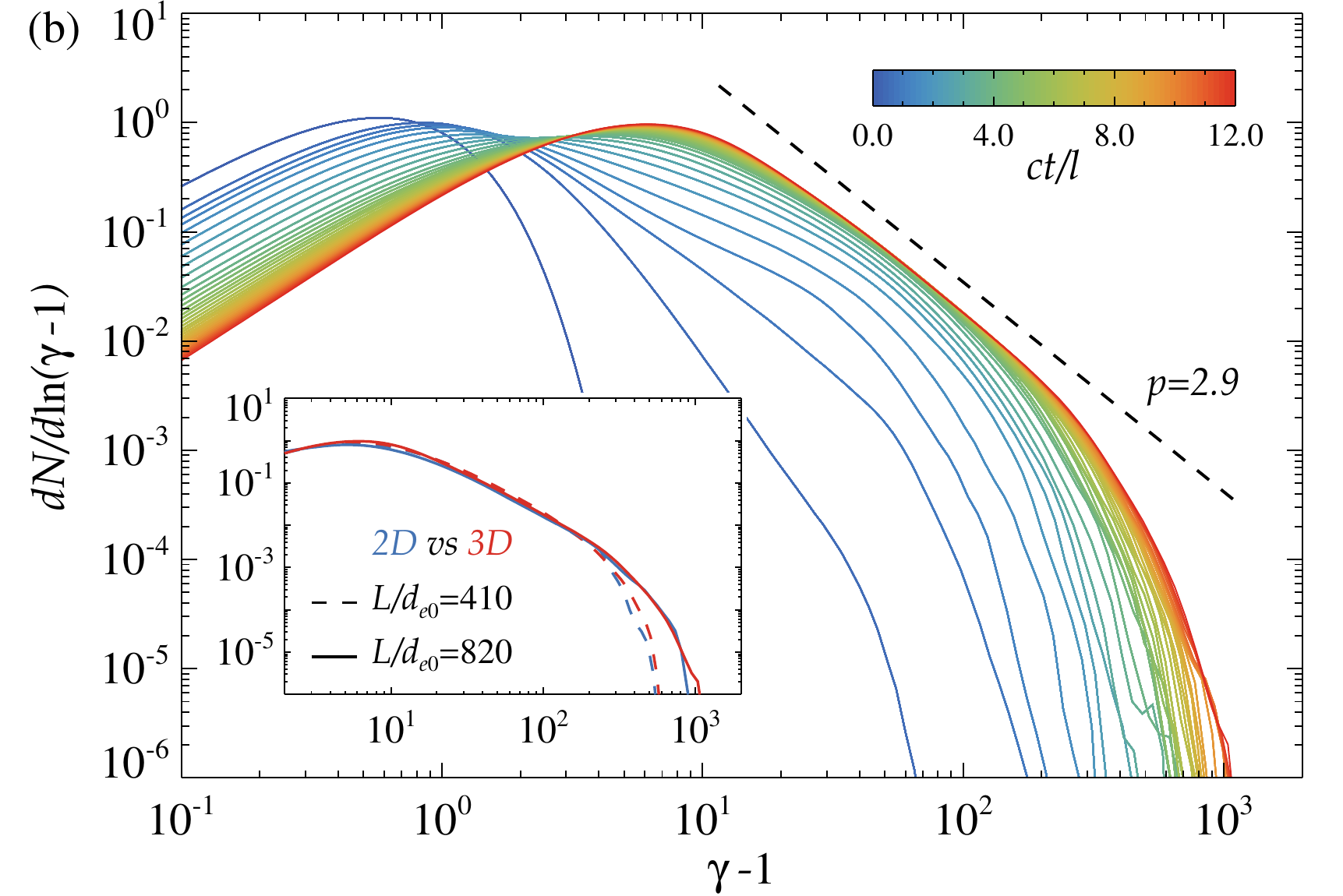}
\end{center}
\caption{Top: Current density $J_z$ at $ct/l=4$ from a 3D simulation with $\sigma_0=10$, $\delta B_{{\rm{rms}}0}/ B_0=1$, $L/d_{e0}=820$, and $l=L/4$, showing the copious presence of current sheets \cite{Movies}. Bottom: Time evolution of the corresponding particle spectrum. The inset shows  for two different box sizes that the time-saturated particle spectra are almost identical between 2D (blue) and 3D (red). }
\label{fig3}
\end{figure}

We have confirmed our main results with large-scale 3D simulations, since several properties of the turbulence itself, as the energy decay rate and the degree of intermittency, are known to be sensitive to dimensionality \cite{biskamp03}. %
Results from our largest 3D simulation, with $L/d_{e0}=820$ and $l=L/4$, are presented in Fig.~\ref{fig3}. The plot of $J_z$ in the fully-developed turbulent state (top) shows the presence of a multitude of current sheets, as found in our 2D setup. The evolution of the particle energy spectrum is presented in Fig.~\ref{fig3}(b). A pronounced nonthermal tail develops, whose power-law slope and high-energy cutoff are remarkably identical to its 2D counterpart (in the inset, we compare the time-saturated spectra of 2D and 3D simulations for two different box sizes, showing that the spectra nearly overlap).

\begin{figure}[]
\begin{center}
\includegraphics[width=8.65cm]{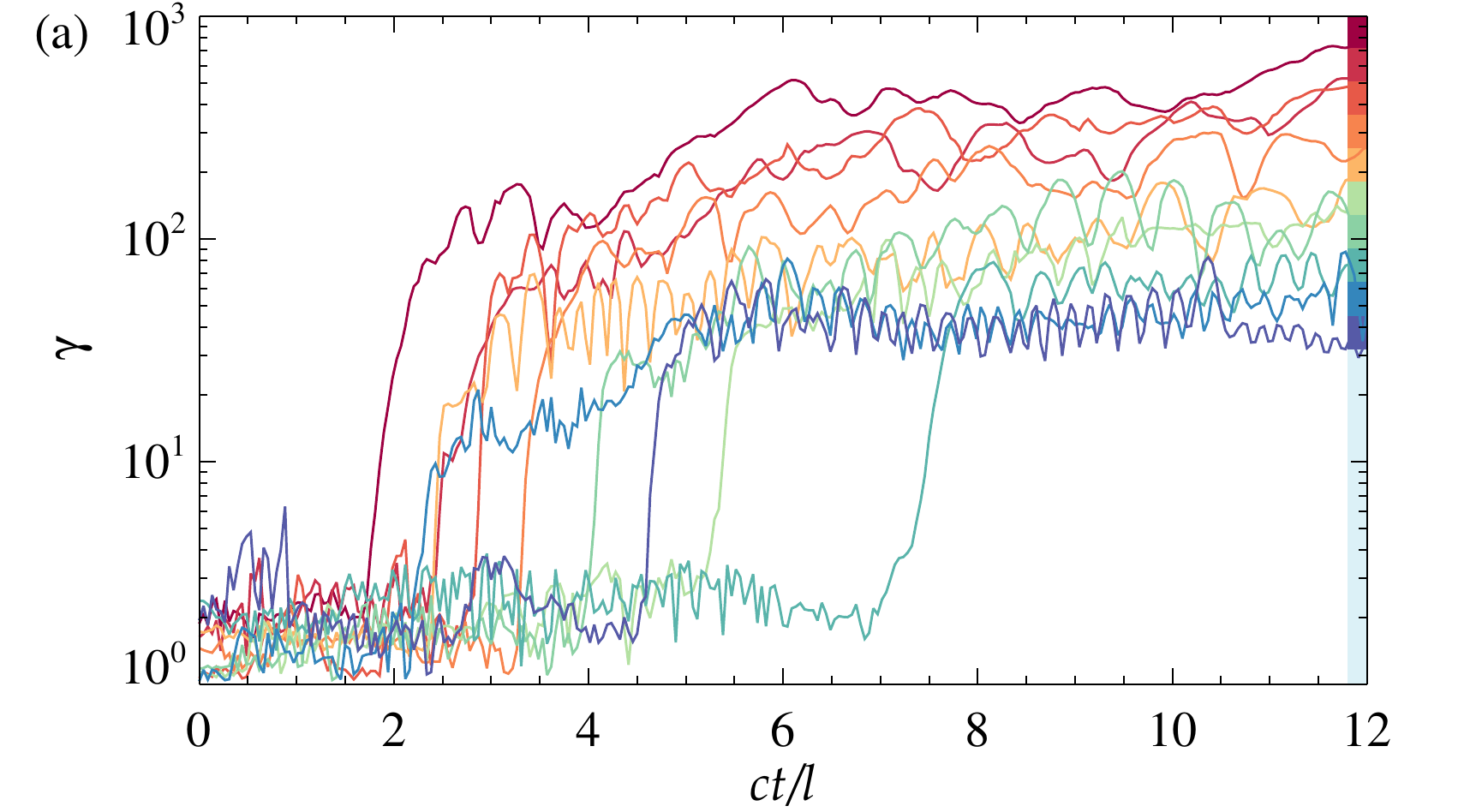}
\includegraphics[width=8.65cm]{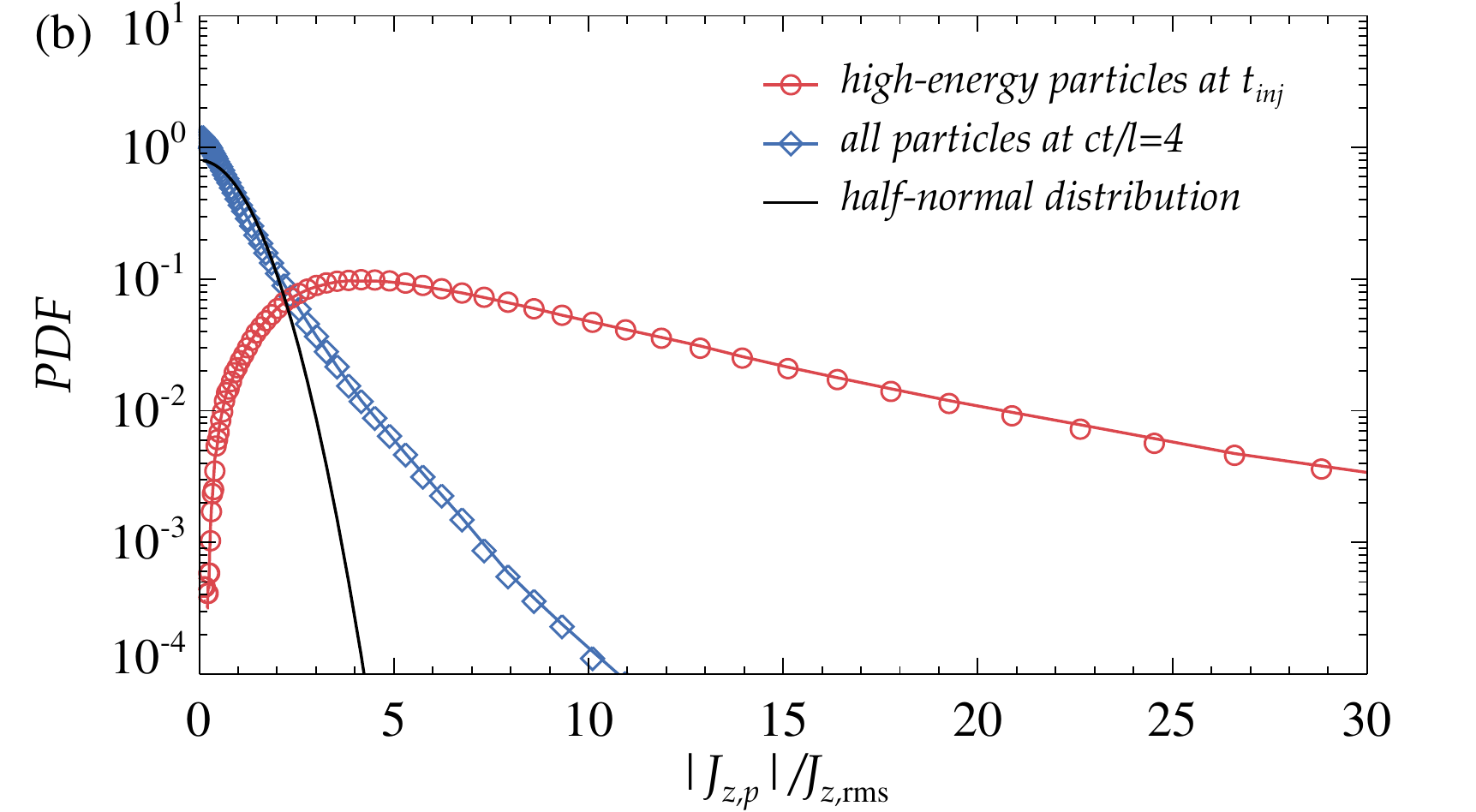}
\hspace*{0.163cm}\includegraphics[width=8.65cm]{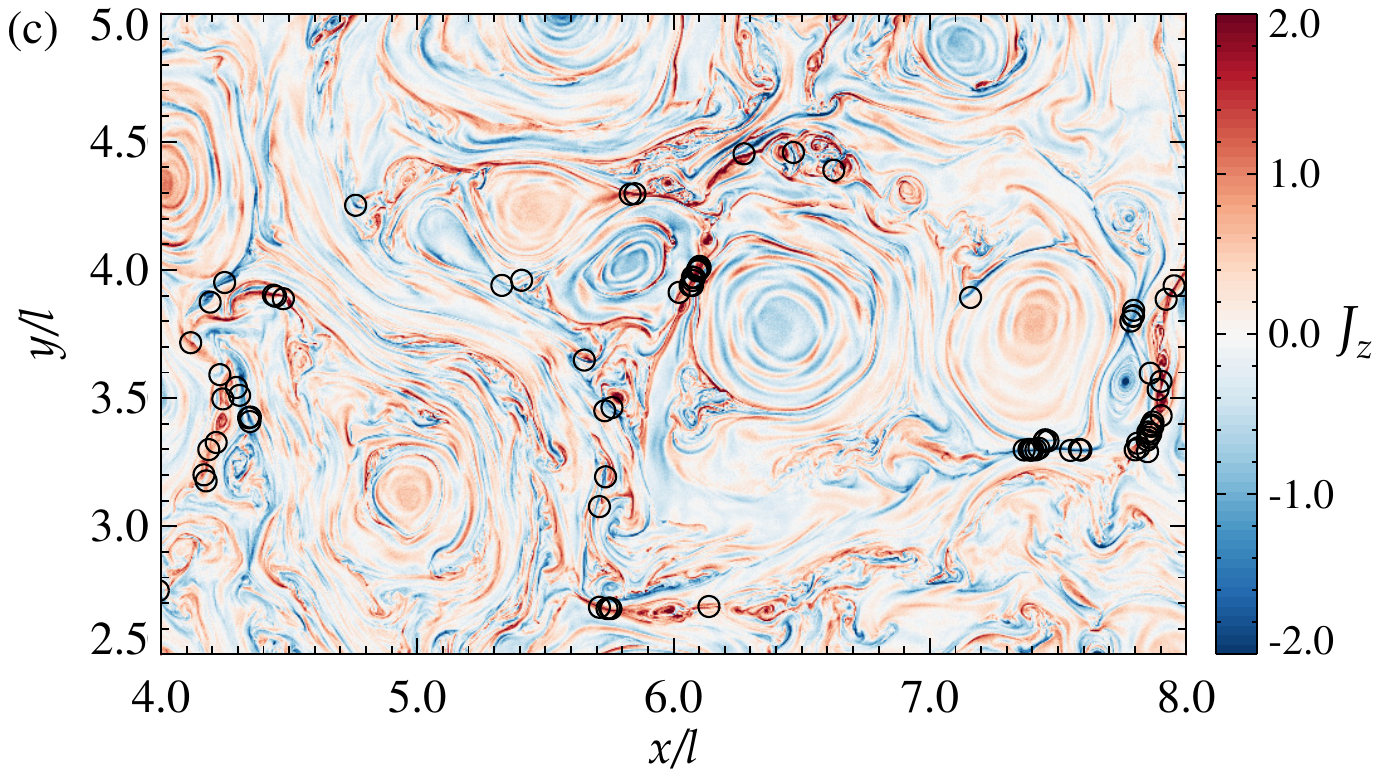}
\end{center}
\caption{Top: Evolution of the Lorentz factor for $10$ representative particles selected to end up in different energy bins at $ct/l=12$ (matching the different colors in the colorbar on the right).
%when the nonthermal spectrum is fully developed. 
Middle: Probability density functions of $|{J_{z,p}}|/{J_{z,{\rm{rms}}}}$ experienced by the injected particles at their $t_{inj}$ (red circles) and by all our tracked particles at $ct/l=4$ (blue diamonds), with a half-normal distribution overplotted as a solid black line. Bottom: Zoom of $J_z$ at $ct/l=4$ with the open circles indicating the positions of the particles that are injected around this time \cite{Movies}.}
\label{fig4}
\end{figure}

{\it Acceleration Mechanisms.---}  
In order to unveil the particle acceleration mechanisms, we have tracked the trajectories of a random sample of $\sim10^6$ particles from a 2D simulation with $\sigma_0=10$, $\delta B_{{\rm{rms}}0}/ B_0=1$, and $L/d_{e0}=1640$. 
In Fig.~\ref{fig4}(a) we show the Lorentz factor evolution of $10$ particles that eventually populate the nonthermal tail (i.e., with $\gamma > 30$ at $ct/l=12$, compare with the cyan line in Fig.~\ref{fig2}(b)). A common feature of these tracks  is the rapid energy increase from $\gamma \sim  \gamma_{th0}$ up to $\gamma \sim 10-100$. Indeed, we have verified that the overwhelming majority ($\sim 97\%$) of the particles belonging to the nonthermail tail (i.e., with $\gamma>30$ at $ct/l=12$) experience in their life such a sudden episode of energy gain.
This event is extracting the particles from the thermal pool and injecting them into the acceleration process (i.e., it controls the physics of particle injection). Inspired by Fig.~\ref{fig4}(a), we identify the injection time $t_{inj}$ as the time when the energy increase rate (averaged over $\Delta t = 45\,d_{e0}/c$) satisfies $\Delta \gamma /\Delta t > {\dot \gamma}_{thr}$, and prior to this time the particle Lorentz factor was $\gamma < 4 \gamma_{th0} \sim 6$. We typically take ${\dot \gamma}_{thr} \simeq 0.01 \sqrt{\sigma_0} \gamma_{th0} \omega_{p0}$, but we have verified that our identification of $t_{inj}$ is nearly the same when varying ${\dot \gamma}_{thr} $ around this value by up to a factor of three. 

Having determined the injection time, we can explore the properties of the electromagnetic fields at the injection location. 
The red circles in Fig.~\ref{fig4}(b) show the  probability density function (PDF) of the electric current density $|{J_{z,p}}|$ experienced by the particles at their injection time (normalized by the root-mean-square $J_{z,{\rm{rms}}}$ in the whole domain at that time). The peak of the PDF is at $|{J_{z,p}}| \sim 4\, {J_{z,{\rm{rms}}}}$, and $\sim95\%$ of the injected particles reside at $|{J_{z,p}}| > 2\, {J_{z,{\rm{rms}}}}$, a threshold that is usually employed to identify current sheets \cite{rappazzo17}. This should be contrasted with the blue diamonds, showing the PDF of the electric current experienced by our tracked particles at $ct/l=4$, regardless of whether they are injected or not. 
%(and if they are, regardless of their injection time). 
As expected, this peaks at zero, and only $\sim9\%$ of particles have $|{J_{z,p}}| > 2\,{J_{z,{\rm{rms}}}}$. The tail of the blue curve at $|{J_{z,p}}| > 2{J_{z,{\rm{rms}}}}$ is due to the intermittent nature of current sheets in turbulence \citep[e.g.,][]{dmitruk04,servidio09,wan12,servidio12,haggerty17,cerri17,dong18}, while for $|{J_{z,p}}| < 2{J_{z,{\rm{rms}}}}$ the blue PDF lies close to a half-normal distribution (solid black line).

In summary, particle injection into the acceleration process occurs at current sheets; more specifically, at reconnecting current sheets. This is illustrated in Fig.~\ref{fig4}(c), where we show $J_z/e n_0 c$ in a subset of the simulation domain at $ct/l=4$. The overplotted open black circles indicate the locations of particles whose $t_{inj}$ is around this time. Clearly, most of the particles participating in the injection episode reside at active reconnection layers, fragmenting into plasmoids.  Despite the small filling fraction of current sheets, the injection efficiency (i.e., the fraction of particles going through the injection phase) is expected to be independent of box size. In fact, the lifetime of a current sheet of characteristic length $l$ is the eddy turnover time $l/\delta {V_{{\rm{rms}}}}$ (here, $\delta {V_{{\rm{rms}}}}$ is the velocity fluctuation amplitude). During this time, reconnection proceeds at a rate $\beta_R c$ and  the current sheet will ``process'' a plasma surface $\sim \beta_R l^2( c/\delta {V_{{\rm{rms}}}})$, i.e., a fixed fraction of the 2D domain (a similar argument holds in 3D).

\begin{figure}[]
\begin{center}
\includegraphics[width=8.65cm]{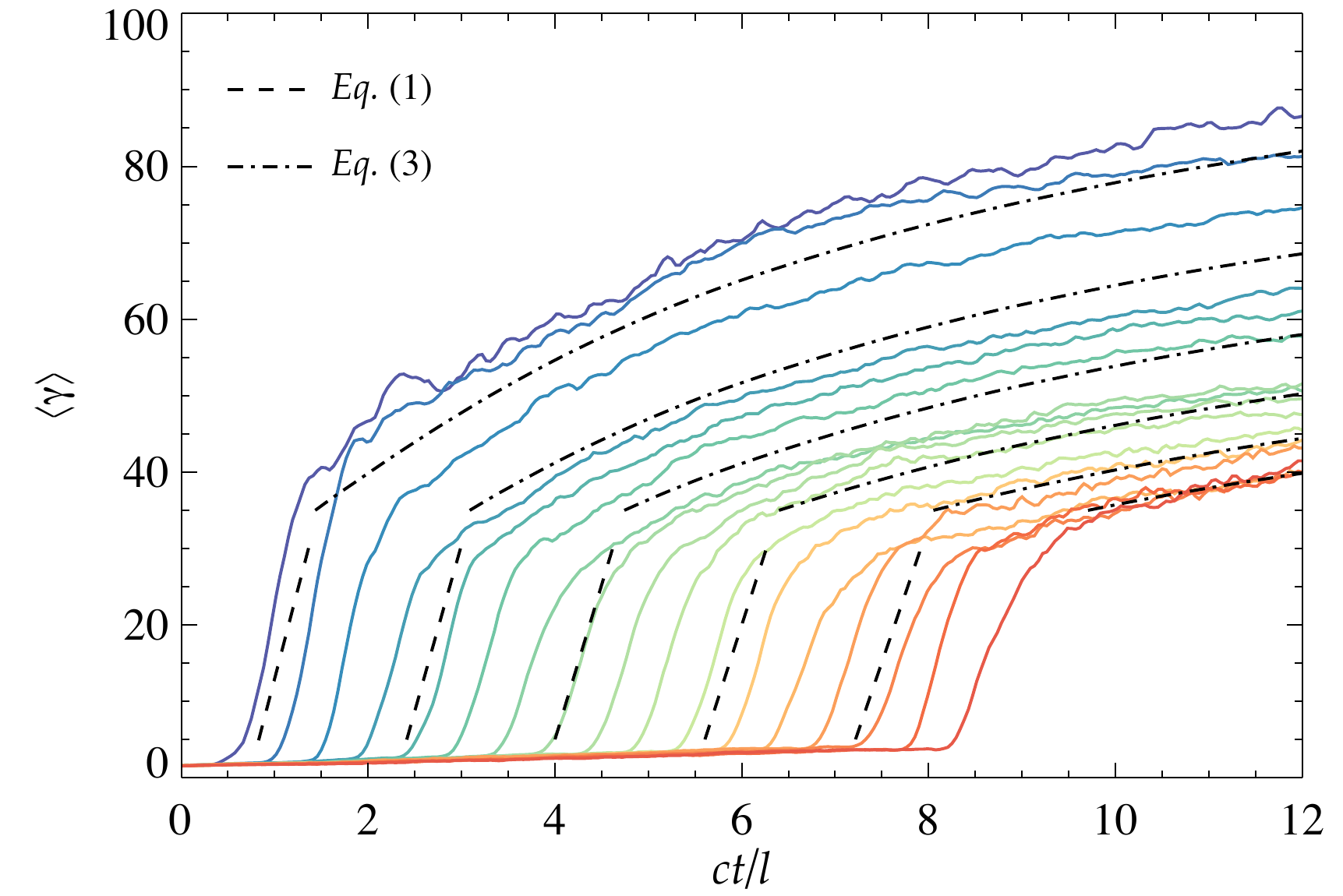}
\end{center}
\caption{Evolution of the mean Lorentz factor of different generations of particles (i.e.,  grouped depending on their injection time $t_{inj}$, in bins of $\Delta t_{inj}=0.48 ct/l$). The initial energy gain, due to the reconnection electric field, can be modeled as in Eq.~(\ref{eq1}) with $\beta_R = 0.05$ (dashed lines), while the subsequent evolution, governed by stochastic interactions with the turbulent fluctuations, follows Eq.~(\ref{eq3}) (dot-dashed line).}
\label{fig5}
\end{figure}

Acceleration by the reconnection electric field \cite{sironi14,guo14,werner16} governs the first phase of particle energization, as shown in Fig.~\ref{fig5}. Here, each colored curve represents the average Lorentz factor of particles having the same injection time $t_{inj}$ (within $\Delta t_{inj}=0.48 ct/l$). The linear growth from $\langle\gamma\rangle\sim 1$ up to $\langle\gamma\rangle\sim30$ (i.e., the injection phase) is powered by field-aligned electric fields, whose magnitude is $|E_{\parallel}|\simeq \beta_R \delta B_{\rm rms}$, via
\begin{equation} \label{eq1}
\frac{d \langle \gamma \rangle}{dt} = \beta_R \frac{\delta B_{\rm rms}}{B_0}\sqrt{\sigma_z (1 + \theta_0/\gamma_{th0})} \,\gamma_{th0} \,\omega_{p0}~.
\end{equation} 
The dashed black lines in Fig.~\ref{fig5} show the predictions of Eq.~(\ref{eq1}) assuming a reconnection rate $\beta_R \simeq 0.05$, as appropriate for relativistic reconnection with guide field comparable to the alternating fields \cite{werner17}.

After the injection phase, the subsequent energy gain (which eventually dominates the overall energization of highly nonthermal particles) is powered by perpendicular electric fields via stochastic scatterings off the turbulent fluctuations. This is a biased random walk in momentum space, which can be modeled with a Fokker-Planck approach \cite{blandford87}, provided that the fractional momentum change in a single scattering is small, as it is the case in our simulations. From the Fokker-Planck equation for relativistic particles,
%(under the usual assumptions of detailed balance and vanishing coefficients for $p \to 0$), 
\begin{equation} \label{eq2}
\frac{d{\left\langle {\gamma} \right\rangle }}{dt}= \frac{1}{\gamma^2} \frac{\partial }{\partial \gamma} \left[ {{\gamma^2} D_{p}} \right] \, ,  \quad D_p  = \frac{1}{3} \frac{\delta V_{\rm rms}^2}{c}\frac{{{\gamma^2}}}{{\lambda_{\rm mfp} (\gamma)}} \,~ ,
\end{equation}
where $D_p$ is the diffusion coefficient in momentum space for a stochastic process akin to the second-order Fermi mechanism, $\delta V_{\rm rms}$ is the  typical velocity of the scatterers (typically ${\delta V_{\rm{rms}}}/c \lesssim 0.3$ in our simulations, which justifies a non-relativistic treatment), and $\lambda_{\rm mfp} (\gamma)$ is the particle  mean-free-path. Since particles are most efficiently scattered by turbulent fluctuations on the scale of their Larmor radius, we assume a Bohm-like scaling for $\lambda_{\rm mfp} (\gamma) = \kappa (c/\omega_L) ( B_0/\delta B_{\rm{rms}})^2$ where $\omega_L = eB_0/\gamma m c$ is the Larmor frequency and $\kappa$ is a dimensionless coefficient. This leads to 
\begin{equation} \label{eq3}
\frac{d \langle \gamma \rangle}{dt} = \kappa^{-1} \frac{{\delta B_{\rm{rms}}^2}}{{B_0^2}} \frac{{\delta V_{\rm rms}^2}}{c^2} \sqrt{\sigma_z (1+ \theta_0/\gamma_{th0})} \,\gamma_{th0} \,  {\omega _{p0}} \, ~.
\end{equation}
Taking the temporal decay of the magnetic and velocity fluctuations directly from our simulation, we obtain for $\kappa=10$ the dot-dashed lines in Fig.~\ref{fig5}, which demonstrate that the Fokker-Planck approach agrees well with our simulation results.

%%%%%%%%%%%%%

%\begin{equation} \label{simplifiedFP}
%\frac{{\partial f}}{{\partial t}} + ({\bm{v}} \cdot {\bm{\nabla}})f  = \frac{1}{p^2} \frac{\partial}{\partial p} \Big[ - F{p^2}f + \frac{\partial }{{\partial p}}\left( {{D_p}{p^2}f} \right) \Big] \, ,
%\end{equation}
%where $f(p,t)$ is the isotropic, homogeneous phase space, $F = {{\left\langle {\Delta p} \right\rangle }}/{\Delta t}$ is the mean momentum gain rate, and $D_p = { \langle { (\Delta p)^2 } \rangle }/{{ 2 \Delta t}}$  is the average rate of momentum dispersion.

{\it Conclusions.---} 
We have demonstrated that turbulence in magnetically-dominated plasmas is a viable mechanism for particle acceleration, since it self-consistently generates nonthermal power-law tails. The power-law slope is harder (near $p\sim 2$) for higher magnetizations and stronger turbulence levels. Thanks to our large domains, we have demonstrated that the power-law slope reaches an asymptotic, system-size-independent value, while the high-energy spectral cutoff increases linearly with system size: this allows to extrapolate our results to the macroscopic scales of astrophysical sources.
The time-saturated particle energy spectrum is remarkably similar in 2D and 3D, suggesting that the same acceleration process operates, regardless of the dimensionality. By following a large sample of particles, we have shown that their energization occurs in two stages: particle injection happens at reconnecting current sheets; this is followed by a phase of stochastic acceleration (which dominates the overall energy gain) where the particles scatter off turbulent fluctuations. Analytical predictions are in agreement with the simulations results, confirming the two-stage nature of the acceleration process. Our results have important implications for the origin of non-thermal particles in high-energy astrophysical sources.

\begin{acknowledgments} 
It is a pleasure to acknowledge fruitful discussions with Mikhail Medvedev, Jonathan Zrake, Vah\'{e} Petrosian, Chuanfei Dong, Yi-Min Huang, Maxim Lyutikov, Vassilis Tsiolis and Loukas Vlahos. This research acknowledges support from DoE DE-SC0016542, NASA Fermi NNX-16AR75G, NASA ATP NNX-17AG21G and NSF ACI-1657507. The simulations were performed on Habanero at Columbia, on NASA (Pleiades) and NERSC (Edison) resources. 
\end{acknowledgments}

\end{document}